\documentclass{elsart}
\usepackage{epsfig}
\usepackage{eurosym}

\begin{document}

\begin{frontmatter}

\title{{\em Ianus}: an Adaptive FPGA Computer }
\vskip -25pt
\author[Ferrara]{F. Belletti},
\author[Bifi]{I. Campos}, 
\author[Zaragoza,Bifi]{A. Cruz}, 
\author[Madrid,Bifi]{L. A. Fern\'andez},
\author[Roma,Bifi]{S. Jim\'enez},
\author[Roma,Bifi]{A. Maiorano}, 
\author[Ferrara]{F. Mantovani},
\author[Roma,INFNroma,Bifi]{E. Marinari},
\author[Madrid,Bifi]{V. Mart\'{\i}n-Mayor}, 
\author[ZaragozaI]{D. Navarro}, 
\author[Madrid,Bifi]{A. Mu\~noz Sudupe},
\author[Zaragoza,Bifi]{S. P\'erez Gaviro}, 
\author[Ferrara]{G. Poli}, 
\author[Badajoz,Bifi]{J. J. Ruiz-Lorenzo},
\author[Ferrara,INFNfer]{F. Schifano}, 
\author[Zaragoza,Bifi]{D. Sciretti}, 
\author[Zaragoza,Bifi]{A. Taranc\'on},
\author[SIC]{P. T\'ellez},
\author[Ferrara,INFNfer]{R. Tripiccione},
\author[Zaragoza,Bifi]{J. L. Velasco}.
\vskip -10pt
\address[Ferrara]{Physics Department, Universit\`a di Ferrara, I-43100 Ferrara
(Italy)}
\address[INFNfer]{INFN, Sezione di Ferrara, I-43100 Ferrara (Italy)}
\address[Bifi]{Instituto de Biocomputaci\'on y  F\'{\i}sica de
Sistemas Complejos (BIFI), Zaragoza}
\address[Zaragoza]{Departamento de F\'{\i}sica Te\'orica,
        Facultad de Ciencias, \\
        Universidad de Zaragoza, 50009 Zaragoza, Spain}
\address[ZaragozaI]{Departamento de Ingenieria Electr\'onica y Comunicaciones,\\
        Universidad de Zaragoza, CPS, Maria de Luna, 1, 50018 Zaragoza, Spain}
\address[Madrid]{Departamento de F\'{\i}sica Te\'orica I, Facultad de CC. 
F\'{\i}sicas, \\
Universidad Complutense de Madrid, 28040 Madrid, SPAIN}
\address[Roma]{Dipartimento di Fisica, Universit\`a di Roma ``La Sapienza''\\
P. A. Moro 2, 00185 Roma, Italy}
\address[INFNroma]{INFN Sezione di Roma}
\address[Badajoz]{Departamento de F\'{\i}sica, Facultad de Ciencias,\\
Universidad de Extremadura, 06071 Badajoz, SPAIN,}
\address[SIC]{Servicio de Instrumentaci\'on Cient\'{\i}fica, Universidad de Zaragoza.}
\vskip -10pt
\begin{abstract}
Dedicated machines designed for specific computational algorithms can
outperform conventional computers by several orders of magnitude. In
this note we describe {\it Ianus}, 
a new generation FPGA based machine 
and its basic features: hardware integration
and wide reprogrammability. Our goal is to build a machine that can
fully exploit the performance potential of new generation FPGA
devices. We also plan a software platform which simplifies its
programming, in order to extend its intended range of application to a
wide class of interesting and computationally demanding problems. The
decision to develop a dedicated processor is a complex one, involving
careful assessment of its performance lead, during its expected
lifetime, over traditional computers, taking into account their
performance increase, as predicted by Moore's law. We discuss this
point in detail.
\end{abstract}
\vskip -15pt

\end{frontmatter}


\section{Introduction}

Dedicated computers have had a long history vis-a-vis several
physics problems demanding huge computing resources, where
conventional computers have not allowed to reach conclusive results.
A non exhaustive list of these problems includes Monte Carlo simulations
of Lattice QCD~\cite{weingarten,APE,APErecent,CHRIST},
numerical studies of large self-gravitating systems \cite{Makino} and
high-resolution numerical solutions of the Navier-Stokes equations in the
turbulent regime \cite{apeNS}.

Statistical mechanics is another area, relevant for our plans, where
dedicated computers have also been heavily
used~\cite{OGIELSKI,HOOGLAND}, starting from the pioneering efforts
dating as far back as the late seventies~\cite{caltech} for the
simulation of the Ising model.  Several dedicated machines have been
developed more recently to study spin-glass systems.  A dedicated
machine, {\it RTN}, built in Zaragoza in 1991~\cite{RTN}, was designed
to perform computationally intensive calculations that make limited
use of floating point arithmetics, such as spin glasses and lattice
Gauge-Higgs model simulations.  RTN was based on transputer
processors: it had 8 identical boards, each with 8 transputers, one
connection chip, and one controller board.  RTN was at the time a very
effective and reliable computational device.  {\it SUE} (Spin
Update Engine) has been the second generation spin-glass
machine~\cite{SUE}.  SUE was completed in the year $2000$, and it has
been used to simulate three dimensional Ising spin glasses.  It
consists of twelve identical boards: each board simulates at the same
time $8$ different systems. The update speed of the whole machine is
$217$ ps/spin (when running at a $48$ MHz clock frequency).  Field
Programmable Gate Arrays (FPGA) made it possible to design a low cost,
high performance, reliable dedicated machine.  The FPGA-based
architecture allows to reprogram the structure of the hardware
connections of the computer: in SUE this feature was used to perform
simple changes, like studying systems of different sizes, or modifying
the updating dynamical scheme, or to change some details of the
Hamiltonian of the system.

SUE was very different from RTN: instead of ``usual'' processors like
transputers (but remember that even transputers were not usual at all
in their extreme optimization toward easy inter-communication) it
adopted FPGA's (of a generation which, looked at from today, seems
prehistoric): these devices give the possibility of ``programming''
the hardware and of configuring the processor at best. In short, they
contain logical gates that can be activated or deactivated or
connected as needed, allowing in this way to select the desired
functionalities.

In this paper we describe yet another generation spin-glass computing
engine, called {\em Ianus} (after the name of the two-faced ancient Roman
God of doors and {\em gates}), whose development is well underway. The
new project has basically two goals: i) developing an extremely
high-performance spin-glass simulation engine, able to update (on
average) one spin in less than $1$ ps and ii) developing a software
infrastructure that drastically reduces the effort to customize an
array of FPGA's to a specific computational algorithm.

This paper is structured as follows: in section $2$ we discuss the
advantages and disadvantages of building dedicated computers instead
of buying commercial clusters, and in section $3$ we describe as a
practical example our computational goals for spin-glass simulations
and the performance of some non-dedicated computers when handling
it. Section $4$ describes the overall architecture of the new machine
and section $5$ discusses the hardware main issues. Section $6$
presents performance estimations, and the paper is ended by some
concluding remarks that also cover longer term plans to make {\em
Ianus} more general purpose.

\section{Advantages and disadvantages of dedicated computing}

The decision to invest time and money in developing a new dedicated
computer has to follow an accurate balancing between real benefits and
development burden; also the possibility of seeing the project
\emph{aging} when compared to actual scientific progress and
technological improvements has to be taken into account.

Optimized computers provide much larger computing power when used in
solving problems they are designed for. This can show up either as a
much better ``price to performance'' ratio, so that the computing
power available within the limits of a given budget is greatly
increased, or as a much higher sheer performance, technically not
achievable at a given point in time on the given problem with
commercial computers. This is not the only way dedicated computers
shorten the time needed to produce scientific results: usually the
development of computer applications for scientific research is
centered in squeezing performance out of very generic hardware. Code
optimization is always a hard and time-eating task and often the gain
is not comparable to the effort. A dedicated machine is built
\emph{around} a specific computational task, and therefore the final
user always accesses its maximum performance.

All computational problems for which dedicated machines have been
proposed have a very large degree of regularity, that typically
translates into heavy use of {\em unusual} mathematical sequences and
into extreme levels of parallelization. Both features are key factors
for performance:

\begin{itemize}
\item 
Traditional computers are optimized for execution of short linear
sequences of code performing irregular memory accesses, and integer
arithmetic or logic manipulations (adds/subtract) of long data words
($32$ or $64$ bits). Recent designs also focus on floating point
arithmetics. Every time the mix of required operations is at variance
with the above list, there is room for dedicated processing. For
instance, the Grape processor is centered around an hardware block
able to compute on the fly the inverse of a square-root.  In Lattice
QCD virtually all processing amounts to manipulation of complex
matrices so that hardware pipelines computing $a \times b + c$ (with
$a, b, c$ complex numbers) are extremely effective.  In our case, as
discussed in details later on, we leverage on the fact that our
typical operations are logical operations on a small number of bits
and that the data-base to be processed is small enough that we store
it within the processor, fully removing the bottle-neck of memory
access.

\item
Traditional computers are not usually designed to be effective for
massive, regular and tightly coupled parallelism. Rather, typical
design goals in commercial parallel systems are extreme flexibility in
possible communication patterns, at the price of interconnection
bandwidth and latency.  Many computationally intensive applications on
the other hand have intrinsically high levels of parallelism, so each
processing node handles a very small subset of the physical
system. This brings inevitably to a large ratio of information
exchanged among nodes over performed arithmetic/logic
operation. Parallel efficiency requires that the time $T_p$ needed by
a processor to handle (e.g., on one iteration) its system subset is
roughly equal to the time $T_c$ needed to exchange needed data with
other nodes. We will see that this requirement is badly violated in
traditional computer systems, while dedicated structures may leverage
on regularity of communication patterns to achieve the goal of $T_p
\simeq T_c$ with limited resources.
\end{itemize}

All this has costs.
\begin{itemize}
\item Large investment in time: a new concept dedicated machine always
requires a long development time, due to designing, building and
testing it.
\item Large investment in man-power: specific high-level knowledge in
electronics, digital design and topics in the target machine's
application fields are needed. Working groups made of engineers,
physicists and biologists have to be formed and coordinated.  In
Universities, the investment in the training of collaborating students
has to be taken into account.
\end{itemize}

\begin{figure}
\includegraphics[angle=270,width=\columnwidth]{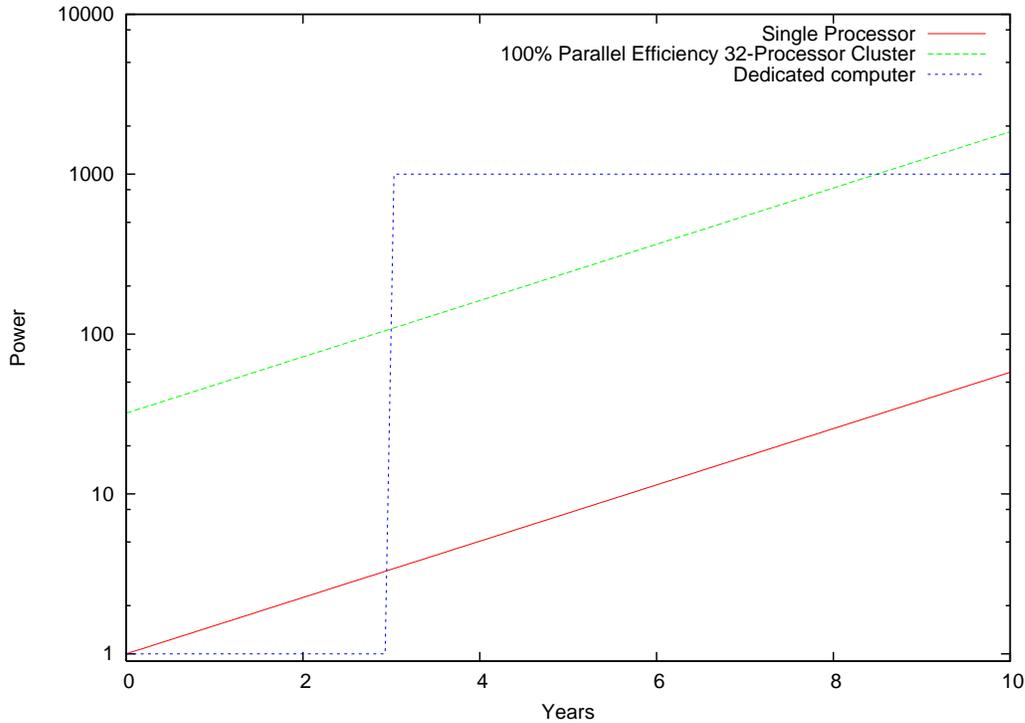}
\caption{Qualitative time analysis of the window of competitiveness of a 
dedicated computer.}
\label{power_evol}
\end{figure}

The relatively long time frame associated to development carries two
additional risks.
\begin{itemize}
\item Possible loss of relevance of the optimized algorithms.  A lot
of care has to be taken when judging what shall be still interesting
to the scientific community in the following years, to avoid that the
specific problems on which the new machine is unbeaten becomes
obsolete before the machine does. We may say that it is the machine
that has to make the problem grow old and not the opposite.
\item Loss of competitiveness in time with respect to commercial
computers.  Computer performance has been growing steadily according
to Moore's law (performance doubles every approximately 18 months),
while the performance of a dedicated machine remains constant, till a
new generation is developed.  Fig. \ref{power_evol} sketches the
situation: one must be sure that the performance gain associated to
the dedicate system, {\em when} it comes into operation is large
enough to offset power improvements of traditional processors working
in parallel with a reasonable parallel factor, and provide a window of
opportunity of at least a few years.
\end{itemize}

In the next sections we analyze these problems with a
specific regard to our application.

\section{Monte Carlo simulations of spin-glass systems}

The first area of application for {\em Ianus} is the accurate and detailed
study of spin-glass systems. Later on we will describe other contexts
in which we expect {\em Ianus} to become a key player. 

We consider here the the Edwards-Anderson spin glass
model~\cite{SG} in three dimensions. We define as usual a
collection of variables $\sigma_i\ ,\ i=1,\dots,N=L^3$ corresponding
to sites of a cubic lattice of size $L$; each variable takes values
$\{-1,+1\}$. A collection of static links $J_{ij}\in \{-1,+1\}\ ,\
i,j=1,\dots,N$ between first neighbor sites determines the properties
of the system, and we are interested in randomly (quenched) chosen ones.  When
studying equilibrium properties of the system, averages on a large
number of realizations of the $\{J\}$ ensemble are appropriate, while
an analysis of the system away from equilibrium requires very long
time histories over a smaller set of $\{J\}$. In the former case,
there is an additional opportunity for trivial parallelization, as
simulations are performed in parallel over different sets of $\{J\}$,
that possibly use the same random numbers. This option is much less
effective in the latter case, whose physical interest is growing with
time, and for which we plan to optimize our new processor.

We discuss here two different algorithms, namely 
the Demon ~\cite{DEMON,Demon_JJ_CL} and the Heat Bath Algorithm ~\cite{HB}.
The main advantage of the former algorithm is
that random number are not needed, so it was widely applied in
earlier dedicated machines. For the latter algorithm, on the other hand,
performance depends strongly on the efficiency of (multiple) random
number generators. 

The energy of the system is defined by 
\begin{equation}
\label{eq:USG}
U = -\sum_{<ij>}\sigma_i J_{ij} \sigma_j \;,
\end{equation}
where $<ij>$ indicates that sums have to be performed over all
pairs of neighboring sites in the physical lattice.
In the Heat Bath algorithm
spins are updated by extracting their values with a probability
given by:
\begin{equation}
\label{eq:PHB}
P(\sigma_i = 1) = \frac{ \e^{\sum_{<ij>} J_{ij} \sigma_j /kT } }
{ \e^{\sum_{<ij>} J_{ij} \sigma_j /kT}
+ \e^{-\sum_{<ij>} J_{ij} \sigma_j/kT }}\;.
\end{equation}
Updates may be done sequentially (one at a time, eventually sweeping
the whole lattice) or in parallel, being careful to respect
balance. The highest level of achievable parallelism is obtained when
all sites are divided into two groups in a checkerboard configurations
and all elements of each group are updated at the same time (use of
checkerboard structure is needed to ensure that detailed balance is
respected during the simulation).

In the Demon algorithm, the system is coupled to a set of energy reservoirs
(``demons'') $e_k\ ,\
k=1,\dots,K$. The energy of this enlarged system
is given by
\begin{equation}
U = -\sum_{<ij>} \sigma_i J_{ij} \sigma_j + \sum_ke_k\;,
\end{equation}
where all the demons may be initially set to zero. The demons cannot lower
their energy below zero and also an upper limit may be assigned. 
Consider the case $K=1$ with a unique reservoir coupled to the system.
Spins are updated sequentially depending on the actual values of the
reservoir: if inverting the value of a spin is energetically convenient, the
demon receives the energy variation and the spin is flipped, otherwise the spin
flips only if the reservoir may supply the needed energy.

Both algorithms can be coded in traditional processors with
reasonable efficiency.

For instance, the Heat-Bath procedure updates a single spin every 10
nanoseconds on a Pentium IV processor at 3.2 GHz by means of a
reasonably programmed multi-spin coding updating routine, in which one
random number is used for each try.  If multiple spins (belonging to
independent ensembles) use the same random number performance increase
by a factor of roughly $3$.

Both algorithms can be parallelized very easily. Typically the
physical lattice is divided in homogeneous region and computations in
each of them are carried out by one computational node; only
boundaries of each partition have to be transported between nodes
after each step of the algorithm.  We have parallelized the Heat-Bath
algorithm on a set of 4 Pentium IV nodes at 3.2 GHz with Gigabit
Ethernet obtaining a parallel efficiency of 42\%, when considering a
lattice of global size $64^3$. The small amount of information to be
transmitted per each complete Monte Carlo lattice update makes latency
overhead very large for each communication. The actual spin update
time (within each node of the cluster) to consider when comparing to our
new dedicated machine is then about 24 nanoseconds.

A properly coded routine for the demon algorithm updates a
spin in less than 4 nanoseconds. The parallel efficiency in this case
is even lower, about 25\%. Sustained parallel update time in this case
is therefore of the order of $15$ nanoseconds in this case.

Very recently, new computer options have become available, in terms of very
massively parallel systems with nearest neighbor connections (Blue Gene/L
\cite{blueGene}) and recently announced processors with extremely powerful
on-board processing resources (like the Clearspeed CSX600 processor
\cite{clearSpeed} or the IBM/\-Sony/\-Toshiba Cell processor \cite{ibmCell}).

In order to estimate the performance of spin-updates on these new
machines, we may build a very simple model, in which we define the
computational load as $\lambda l^3$ ($l$ is the linear size of the
lattice handled by each processor), and the information-exchange
between neighboring nodes $ 6 \iota l^2$. For a balanced spin-update
engine, we will have
\begin{equation}
\label{eq:balance1}
T_p = \lambda l^3 / F  \simeq T_c = 6 \iota l^2  / B\;,
\end{equation}
where we have introduced average sustained processing power $F$ and
average sustained interconnection bandwidth $B$ (both quantities can
be very conveniently written in terms of operations (or bits
exchanged) per processor clock cycle). Considering for definiteness
the heat-bath algorithm, we have $\lambda \simeq 10$ (in $\lambda$ we
have lumped a few short-word adds, the computation of one random
number and access to one entry of a small look-up table), and $\iota =
1$. Comparing with the performance figures for the Pentium IV cluster,
we have effective values of $F \simeq 0.3 $, and $B \simeq 0.01$.

For the new machines, we use published data for $B$ and an approximate
estimate for $F$ based on the assumption that the (clock-frequency
normalized) performance of each integer processing unit available on
these new processors is roughly the same as that available on a Pentium 4
(probably an upper limit, since these processors are all geared to
floating point performance). We list relevant properties in table
\ref{tab:procs}, where we also estimate the update time per site and
the smallest value of $l$, $l_\mathrm{min}$, for which performance is not limited by
communications, that we derive from Eq. (\ref{eq:balance1}).

\begin{table}
\begin{center}
\begin{tabular}{|c||c|c|c|}
\hline
-                & Blue Gene      & Cell            & CSX600\\ \hline
Clock (GHZ)      & 0.7            & 3.2             & 0.25\\
$F $(ops/cycle)  & 0.3            & 2.4             & 32\\
$B $(bit/cycle)  & 12             & 192             & 256\\
$l_\mathrm{min}$ & $ \ll 1$       & $ \ll 1$        & $ \ll 1$ \\
Update Time   & $\simeq 45$ ns & $\simeq 1.25$ ns & $\simeq 1.33$ ns\\
\hline
\end{tabular}
\label{tab:procs}
\caption{Architectural parameters for Blue Gene/L and for the IBM-Cell, and
ClearSpeed CSX600 processors and their corresponding 
single-processor estimated spin-update time.}
\end{center}
\end{table}

Note that these processors and systems are strongly optimized for
floating-point calculations, not particularly useful in our
context. This is worsened by the fact that fast floating point is
single precision for Blue-Gene and the Cell processors (the latter
also has less accurate rounding circuits) so its use for random number
generation is questionable. For this reason sustainable $F$ values are
smaller than would be achievable in floating point intensive
computation. Inspection of table \ref{tab:procs} shows that the
update-speed of one Blue-Gene/L node is smaller than on a conventional
processor, while both the Cell and CSX600 processors promise a boost
of almost one order of magnitude more than traditional high-end
microprocessors.  The real advantage potentially offered by all these
new architectures however is that all the intrinsic parallelism of the
underlying problem can be exploited without performance penalties.
 
\section{An overview of {\em Ianus}}

{\em Ianus} is a massively parallel machine, optimized for the
simulation of spin-glass systems and based on latest generation FPGA
components.

In recent years, FPGA's have increased dramatically their processing
speed and, above all, the logic complexity that they can
incorporate. At present, logical systems of the order of millions of
gates can be built inside just one FPGA device. At the same time,
memory blocks of the order of millions of bits are built into high-end
FPGA's.  As already stressed, we expect to make much better than
conventional processors in application domain, leveraging on three
main cornerstones:
\begin{itemize}
\item logic operations associated to spin-glass simulations differ
from the arithmetic operations in which traditional computers focus.
\item our data-base can be fully contained by memories inside the processor.
As a consequence an extremely high bandwidth is available to feed processing
blocks.
\item FPGA's have very large resources for off-chip communication that
we exploit for processor to processor data-transfer, with simple protocols
that help reduce latency.
\end{itemize}

At the same time, FPGA's should help reduce development efforts,
enlarging the window of opportunity for our machine, and allow, at a
later stage, efficient tailoring of the developed hardware in
different application areas.

In short, we are developing a {\em bi-programmable} machine, that we
call {\em Ianus}. {\em Ianus} is based on the following main
components:
\begin{itemize}
\item A hardware layer, based on a module containing several high-end FPGA's.
\item A communication fabric that allows high bandwidth data exchange
among the FPGA's and connection of the hardware module with a
traditional host computer.
\item A software layer that allows to load a specific simulation model onto the
hardware layer, set appropriate simulation parameters, run actual simulations
and collect results.
\item At a later stage, we plan to add a model-development layer that
allows to develop an ad-hoc FPGA based simulation engine for a large
and growing class of interesting models with reasonable effort (we
refer to this layer with the term bi-programmability).
\end{itemize}

The natural focus of {\em Ianus} will be toward discrete variables. Binary
variables are the easiest example (and at first glance they can be
seen as occupation variables, or as magnetic spins, or as boolean
truth statements), but we plan and demand that {\em Ianus} should be as
effective when dealing with $10$ or $50$ state variables. Typical
fields we believe will be able to use {\em Ianus} effectively are:

\begin{itemize}

\item Lattice models for physical systems.

\item Slow dynamics on simplified models: that is very important when
      discussing the former issue but is also very relevant for the
      next two issues.

\item Biological issues: Go models, statistical mechanics of DNA and
      RNA, docking.

\item Optimization and Cryptography problems.

\end{itemize}

\section{{\em Ianus} hardware}
The hardware architecture of {\em Ianus} has already been developed: the
basic computational block will contain 16 high-end FPGA chips (e.g.,
of the Altera Stratix or Xilinx Virtex families) that we call SP, (for
Simulation Processors).  We believe that using a powerful
state-of-the-art core for the SP is an important feature of the
machine we have in mind.  Notice that each SP processor will be able
to deal with of the order of $10^5$ logical elements at the same time (at
least a $32^3$ lattice of Boolean variables when using binary link
coupling). 

The SP processors will be arranged and connected in a two dimensional array of
$4\times 4$ processors: periodic boundary conditions will be enforced by
hardware.  A seventeenth processor will be used as an Input/Output Processor
(IOP) for communicating with other boards. It will play the role of a cross-bar
switch: each of the $16$ SP will have a (fast) link to the IOP. The IOP will
perform two main functions, by allowing effective long range SP to SP
communication and by acting as interface between the processing board and the
host computer, a traditional Linux-based system.

In the long term,
this hardware structure will be enabled by a growing set of FPGA-optimized
building blocks (random number generators, Metropolis update engines, 
address-generators ...) to be used to make the SP's the dedicated/optimized
processors for each target application.
We  also plan a user friendly framework interface that helps to assemble
the wished set of computational functionalities, as well as
a run-time support system that controls the configuration of the SP's and
the actual simulation steps.

\section{Performance estimates for {\em Ianus}}
We have already developed and tested preliminary versions of the FPGA
based implementation of both the Demon and Heat-Bath algorithms. They
are being tested on simple test boards using recent FPGA devices
(Altera EP1S60 and Xilinx Virtex 4 LX25). For the final version of the
machine we are considering either devices of the Altera Stratix EP2 or
Xilinx Virtex 4 series. These more recent components have roughly up
to 100\% more available gates (and memory elements) and their speed is
some 20\% 30\% higher. Our preliminary results are therefore lower
limits on achievable performance.

We have developed and tested a simulating engine for the demon algorithm that
tries to extract all possible parallelism. Our implementation of the algorithm
is as follows: inside each system every site may be labeled as \emph{black}
or \emph{white} following a checkerboard scheme, and each of the two subset may
be updated in parallel, all neighbors of a black site being white ones. We
define $K=N/2$ demon reservoirs, the first one of them  coupled to the first
black spin and the first white one and so on; then we proceed with parallel
update within each set, alternating the simultaneous update of $N/2$ black
spins and $N/2$ white ones. When simulating spin glasses one usually defines
two identical replicas of the system, with same fixed $\left[J_{ij}\right]$
coupling configuration and different independent initial spin configurations;
this leads to an immediate improvement of the algorithm: we put all black
spins of replica 1 and all white spins of replica 2 in the respective
positions of an ``artificial'' lattice system, say the \emph{P lattice}, while
in the \emph{Q lattice} we put black spins of replica 2 together with white
spins of replica 1. This way, each spin in P has its neighborhood in Q and
vice versa. Also we double the number of demons to $K=N$. This way we can
update in parallel a whole lattice of $N$ spins, the P or the Q one in turn.

Preliminary tests show that we need only a 5 ns clock cycle to perform the
simultaneous update of up to 1000 spins, that corresponds to a lattice
size $L=10$. In doing this, we only made use of flip-flop registers to store
variables and the connectivity has been ``hard-wired''; an ``update engine''
module is defined (taking as input a spin variable, its demon and neighborhood
and outputting updated spin and demon) and the updating logic is replicated
once for each site of the lattice; each variable register is directly
connected to the appropriate ports of the appropriate engines. This approach
is very efficient but pays the cost of being logic consuming: the
``finiteness'' of an FPGA strongly limits the maximum size $L$. This way we
can simulate up to $L=10$ on  an Altera Stratix EP1S60 and up to $L=6$ on a
Xilinx Virtex 4 LX25.

In order to improve the size on these preliminary tests, we can hardwire
connections only on one plane of the cubic lattice, implementing $K=L^2$
demons and updating engines, saving logic to store systems of size up to $L=10$
on the relatively small Xilinx Virtex4 xc4vlx25 device and up to $L=14$ on the
Altera Stratix EP1S60. Even if this way we need $L$ clock cycles to update the
whole lattice, we can update a spin every $0.1$.
Also a wise use of memory blocks inside FPGA's guarantees raising sizes by a
factor 3 and more, with a small cost in speed.

On the final machine, we can easily subdivide a $64\times 64\times 64$ system
among the 16 SP's of the dedicated machine board in slices of $4\times
64\times 64$. Cautiously, we update a $4\times 1\times 64$ slice at a time,
taking no more than 64 cycles to update in parallel all the system (a quarter
of million spins!). Exploiting low latencies, the algorithm may be tailored
to send computed boundary data (two faces of $64\times 64$ data bits)  while
the sequential update is running inside each SP. $B = 128$ 
would provide a sustained speed of $16 \times
4\times 64 = 4096$ spins per clock cycle, with a parallel efficiency of 100\%.
With a clock cycle of 5 nanoseconds, it corresponds to less than $1.3$
picoseconds per spin update.

For the Heat Bath algorithm we need a random number generator for each site
that we want to update in parallel. 
We have implemented an FPGA version that includes 128 update engines,
each equipped with its own random number generator (based on the shift
register algorithm of \cite{randomAlgo}, with 32-bit words).
At present, for test purposes, we use a lattice of $32^3$ sites completely
contained within the processor. The system updates one two-dimensional plane
at a time. Using 128 update-engines in parallel, each plane is updated in
8 clock cycles, then the following planes are processed.

If we consider the same machine structure as above, and the same physical
lattice, we may split the physical lattice in several ways. One possible
solution, that minimizes bandwidth requirement has each processor handling
a sublattice of $16 \times 16 \times 64$ lattice points. Our processors
update $128 \times 16$ spin per clock cycles, corresponding to an update time
of approximately $2.5$ ps. This performance is sustained by a bandwidth
as small as 32 bits per clock cycle.

Extrapolating our preliminary results, we can conclude that {\em
Ianus} (with, e.g., 16 processing boards) will be equivalent to a
Cluster of $ \simeq 5 \times 10^4$ Pentium IV processors or to a
partition of $\simeq 3 \times 10^5$ Blue-Gene/L processors, or to an assembly
of $ \ge 8000$ Cell or Clearspeed processors.

At present we are controlling that the Random Generators we will use
are correct, and also that the parallel updating scheme does not
introduce larger correlations between configurations: eventually the
plane-sequential version of the implementation should be safer.  At
the same time we are finalizing the FPGA design and starting to design
the actual hardware infrastructure and the software structure needed
to interface to traditional Linux-based host computers.

\section{Conclusions}
Probably what is more important to stress here is that {\em Ianus}
wants to be a machine that can solve a large number of problems. We
believe that even models defined on random graphs, or where variables
can take a finite number of values, will be very well suited for {\em Ianus}:
problems like the study of Go models for protein folding and, more in
general, biological issues, optimization issues and more will fit very
well this pattern.

We already showed that it is possible to set up a computing device with an
absolutely favorable performances-to-costs ratio, which is likely to
be long-lived in the field of Ising spin glasses simulations. Keeping in
mind that several topics in the field of complex systems are strictly
related to the physics of Ising spin systems (among the classes of
problems cited above, for example, error correction codes and the
satisfiability of Boolean formulae), and that disposable technology is
giving us many more possibilities with respect to the previous (and
fruitful) SUE experience, we are confident in {\em Ianus} turning out
as a ``dedicated'' and ``multi-purpose'' machine at the same time:
another two-faced aspect justifying the choice of its name.

{\bf Acknowledgments}

We wish to thank Piero Vicini, Alessandro Lonardo, Davide Rossetti and
Sergio De Luca for very useful discussions. We also thank Liliana
Arrachea, Pierpaolo Bruscolini, Kenneth Dawson, Giacomo Marchiori,
Yamir Moreno Vega, Giorgio Parisi, Ilenia Pedron and Federico
Ricci-Tersenghi for sharing with us interesting ideas.  This work has
been partially supported by DGA, MEC (BFM2003-C08532-C03, FIS04-5073-C04, FISES2004-01399, FPA04-2602) and the European
Community's Human Potential Programme, contracts
HPRN-CT-2002-00307 (DYGLAGEMEM) and HPRN-CT-2002-00319 (STIPCO).

\end{document}